\documentclass[aip,jcp,twocolumn,graphicx,amsfonts,amssymb]{revtex4}

\renewcommand{\onlinecite}[1]{\citenum{#1}}

\usepackage{graphicx}

\newcommand{\R}{\mathbf{r}}

\begin{document}

\title{Orbital--dependent second--order scaled--opposite--spin correlation
functionals in the optimized effective potential method} 

\author{Ireneusz Grabowski}
\affiliation{Institute of Physics, Faculty of Physics,
Astronomy and Informatics, Nicolaus Copernicus University, Grudziadzka 5,
87-100 Torun, Poland}
\author{ Eduardo Fabiano}
\affiliation{National Nanotechnology Laboratory, Istituto
Nanoscienze--CNR, Via per Arnesano, I-73100 Lecce, Italy.}
\author{Andrew M. Teale}
\affiliation{School of Chemistry, University of Nottingham,
University Park, Nottingham, NG7 2RD, UK.}
\affiliation{Centre for Theoretical and
Computational
Chemistry, Department of Chemistry, University of Oslo, P.O. Box 1033
Blindern, N-0315 Oslo, Norway.}
\author{Szymon \'Smiga}
\affiliation{Institute of Physics, Faculty of Physics,
Astronomy and Informatics, Nicolaus Copernicus University, Grudziadzka 5,
87-100 Torun, Poland}
\author{Adam Buksztel}
\affiliation{Institute of Physics, Faculty of Physics,
Astronomy and Informatics, Nicolaus Copernicus University, Grudziadzka 5,
87-100 Torun, Poland}
\author{Fabio Della Sala}
\affiliation{National Nanotechnology Laboratory, Istituto
Nanoscienze--CNR, Via per Arnesano, I-73100 Lecce, Italy.}
\affiliation{Center for Biomolecular Nanotechnologies
@UNILE, Istituto Italiano di Tecnologia (IIT), Via Barsanti, 73010 Arnesano
(LE), Italy.} 

\date{\today}

\begin{abstract}
The performance of correlated optimized effective potential (OEP) functionals
based on the spin--resolved second--order correlation energy is analyzed. 
The relative importance of singly-- and
doubly-- excited contributions as well as the effect of scaling the same-- and
opposite-- spin components is investigated in detail comparing OEP results
with Kohn--Sham (KS) quantities determined via an inversion procedure
using accurate ab initio electronic densities.
Special attention is dedicated in particular to the recently proposed
scaled--opposite--spin OEP functional [I. Grabowski, E. Fabiano and F. Della
Sala, \emph{Phys. Rev. B}, \textbf{87}, 075103, (2013)] which
is the most advantageous from a computational point of view.
We find that for high accuracy a careful, system dependent, selection of the 
scaling coefficient is required. We analyze several size-extensive approaches 
for this selection. Finally, we find that a composite approach,
named OEP2-SOSh, based on a post-SCF rescaling of the correlation energy
can yield high accuracy for many properties, being comparable with the most 
accurate OEP procedures previously reported in the literature but at
substantially reduced computational effort.
\end{abstract}

\maketitle

\section{Introduction}
In recent years, {\it ab initio}  correlation energy functionals 
for use in
density-functional theory (DFT) 
have raised considerable interest, since they provide a systematic way to 
overcome the limitations of conventional (i.e. local or semi--local)
density--dependent 
approximations \cite{dftbook3}, such as the presence of  self--interaction error,
qualitatively incorrect correlation
potentials \cite{umrigar:1994:EXACT,grabowski:2011:jcp},  description of dispersion 
interactions \cite{furche11,galli11} and the Kohn--Sham (KS) occupied-virtual energy--gaps~\cite{grun06,ls2}. 
The development of such functionals has followed different paths including 
the adiabatic--connection fluctuation--dissipation (ACFD)
theorem~\cite{lang75,lang77}, G\"orling--Levy Perturbation Theory (GLPT) \cite{gorling:1993:PT}, many--body  perturbation theory (MBPT)
\cite{kutz09,grabowski:2002:OEPP2,ivanov:2003:P2,schweigert:2006:pt2}, and
the idea of {\it ab
initio} DFT \cite{bartlett:2005:abinit2}, in which the density condition
\cite{gorling:1995:IJQCS} together with coupled--cluster (CC) methodology is 
employed. 
In all cases the resulting correlation functionals
depend explicitly on all the KS orbitals 
and eigenvalues (i.e. both occupied and virtual). Thus, for a full self--consistent--field (SCF) solution 
of the KS equations, the optimized effective potential (OEP) method 
\cite{goe05,kummelrmp08,lhf2,engel11book} must be employed.

The OEP method is nowadays widely established for exact exchange (EXX or OEPx) 
in calculations  on  both molecular 
\cite{gorling:1999:OEP,ivanov:1999:OEP,yang:2002:OEP,hesselmanexx07,heatonburgess07,filatov10} 
and solid--state systems \cite{Stadele97,grun06,sharma07,blugel11}.
Hence, the OEPx method and its approximations \cite{lhf1,lhf2,ceda,elp,staroverov2} are gaining 
popularity,  
especially because they remove the one--electron self--interaction error (SIE) and 
strongly improve the KS eigenvalue spectrum as compared to conventional density--dependent
functionals \cite{goe05,kummelrmp08,lhf2,engel11book}. 

In contrast, correlated OEP calculations are much 
less common in the literature and they are still an open research topic for 
development and testing. In particular, correlated calculations based on the ACFD approach, 
starting from the Random--Phase--Approximation (RPA) level,
are usually only performed in a post--SCF fashion, i.e. using orbitals and 
eigenvalues from conventional KS calculations employing a semilocal functional 
 \cite{furche08,kresse09,xen11,hess11,furche12,hutter13} or the
OEPx \cite{goe11,goe10,hess11} method.
In fact, a stable and efficient full--SCF OEP--RPA solution is difficult to achieve 
\cite{hellgren07,hellgren12,verma:044105} and only very recently 
an approach for large systems has been presented \cite{goe13}.
Moreover, the direct RPA  i) shows large 
inaccuracies for thermochemistry, so that different extensions have 
been 
presented
in  recent years
\cite{paier10,xen11,furche13,sosex,ren13}, but all lacking
a corresponding correlation potential;
ii) has a computational cost about one order of magnitude 
larger than the one for the second--order correlation \cite{hutter13};
iii) requires larger basis sets than second--order correlation 
\cite{kresse12,furche12basis,fabiano12}.

More successful implementations
of full SCF correlated OEP calculations
have been obtained in the context of perturbation theory,
in most cases restricted to the use of
second--order correlation.
Thus, as a common practice
correlated OEP calculations are based on the
second--order GLPT (OEP--GL2), since this leads to a
reliable and physically sound,
although overestimated
\cite{bonetti01,grabowski:2002:OEPP2,bartlett:2005:abinit2,mori-sanchez:2005:oeppt2,schweigert:2006:pt2}, 
description of
correlation effects. Nevertheless, because GLPT2 
is unbounded from below, a variational collapse is possible
for unphysical exchange--correlation potentials 
\cite{rohr-baerends:2006:oephe,engel:2005:oeppt2}. 
These potentials do not correspond to KS solutions 
and can be in general easily excluded by minimal regularization 
(such as the truncated SVD approach used in the present work). 
Nonetheless for special cases numerical instabilities can persist.

To improve stability and accuracy new
approaches have been proposed. One of these
is based on a modified MBPT(2) functional with  
a semi--canonical transformation
of the Hamiltonian (OEP2--sc)
\cite{bartlett:2005:abinit2,schweigert:2006:pt2,grabowski:2007:ccpt2}.
As an alternative, recently
a new method (OEP2--SOS) \cite{sos_oep2:prb}
has been proposed in this context, based on
the spin--resolved second--order correlation energy expression \cite{grimme03},
and especially on the scaled--opposite--spin (SOS) variant thereof.
This method takes advantage of the improved accuracy demonstrated for
spin--component--scaled (SCS) and SOS second--order M{\o}ller--Plesset (MP2)
calculations \cite{WCMS12,grimme03} and provides an accurate
OEP correlation functional that largely outperforms the OEP--GL2 approach
\cite{sos_oep2:prb}.
Moreover, the OEP2--SOS method demonstrated a clear advantage over the
OEP--GL2 method, converging also in difficult cases 
(e.g., the Beryllium atom).  
In fact the OEP2-SOS
method employs a scaled (smaller) correlation potential as compared to
OEP-GL2, and thus the
former yields larger energy-gaps and it is less prone to variational
collapse than the latter\cite{sos_oep2:prb}. 
 Finally, because only the
opposite--spin correlation is involved in its formulation,
the OEP2--SOS functional has a favourable
computational cost with respect to other second--order approaches
($\mathcal{O}(N^4)$ vs. $\mathcal{O}(N^5)$).

The OEP2--SOS method is thus a promising approach for {\it ab initio}
correlated DFT calculations. Nevertheless, a systematic and detailed test of 
its performance is still lacking, since in Ref. \onlinecite{sos_oep2:prb}
only a few systems were considered in the test set. Moreover,
no investigation was performed to
inspect the role of the different components of the spin--resolved
second--order correlation (namely  the opposite--spin (OS), same--spin (SS), 
and singly excited (SE) contributions) and determine optimal scaling 
coefficients for different properties. 
In this paper we conduct a thorough investigation
of the SCS and SOS second--order OEP correlation,
by analysing the specific contribution of the different
components of the spin--resolved second--order OEP correlation
and the overall performance of several variants of the method.
To perform this analysis we compare the correlation energies obtained with 
accurate coupled-cluster singles-doubles with perturbative triples [CCSD(T)]
values. In addition, we compare the resulting KS orbital energies and
electronic densities with accurate KS values determined using an inversion
procedure based on the CCSD(T) electronic densities. 

\section{Theory}\label{Sec:Theooepc}
For a given orbital-- and eigenvalue--dependent exchange--correlation (XC) energy functional ($E_{\mathrm{xc}}$)  the 
OEP equation for the XC potential reads
\cite{sharp:1953:OEP,talman:1976:OEP,krieger:1990:OEP,gorling:1995:IJQCS,grabowski:2002:OEPP2,kummel:2008:oep,lhf2,engel:2003:3GDFT}
\begin{equation}\label{e1}
\int  X_\sigma(\R,\R')  v_{\mathrm{xc},\sigma}^{\mathrm{OEP}}({\R'}) \mathrm{d}\R'=\Lambda_{\mathrm{xc},\sigma}(\R) \,  ,
\end{equation}
which is an integral equation  (Fredholm of the  first kind) with the inhomogeneity
given by
\begin{eqnarray}\label{e2}
\Lambda_{\mathrm{xc},\sigma}(\R)&=& \sum_{p}\bigg\{  \int \phi_{p\sigma}(\R)
 \sum_{q\ne p} \frac{\phi_{q\sigma}(\R)\phi_{q\sigma}(\R')}
{\epsilon_{p\sigma}-\epsilon_{q\sigma}}
\frac{\delta E_{\mathrm{xc}}}{\delta\phi_{p\sigma}(\R')} \mathrm{d}\R' \nonumber \\
&&+ \frac{\delta E_{\mathrm{xc}}}{\delta\epsilon_{p\sigma}}|\phi_{p\sigma}(\R)|^2 \bigg \}
\end{eqnarray}
where  the index $\sigma$ labels the spin (throughout this work we 
denote the spin indicies $\sigma,\tau\ldots$),
$X_\sigma^{-1}$ is the inverse of the static KS linear 
response function
\begin{equation}
X_\sigma(\R',\R) = 2\sum_{ia}\frac{\phi_{i\sigma}(\R')\phi_{a\sigma}(\R')\phi_{a\sigma}(\R)\phi_{i\sigma}(\R)}{\epsilon_{i\sigma}-\epsilon_{a\sigma}}\ ,
\end{equation}
and $\phi_{p\sigma}$ and $\epsilon_{p\sigma}$ denote the KS orbitals and eigenvalues, 
respectively, determined by the KS equations
\begin{equation}
\left[-\frac{1}{2}\nabla^2
+v_{\mathrm{ext}}({\bf r})
+v_{\mathrm{J}}({\bf r})
+v_{\mathrm{xc},\sigma}^{\mathrm{OEP}}({\bf
r})\right]\phi_{p\sigma}({\bf r})
=\epsilon_{p\sigma}\phi_{p\sigma}({\bf r})\ ,
\end{equation}
with $v_{\mathrm{ext}}({\bf r})$ and $v_{\mathrm{J}}({\bf r})$  the external (nuclear) and the Coulomb potentials, respectively.
In all equations we use the convention
that $i,j,k$ label occupied KS orbitals, $a,b,c$ label virtual ones,
while the indexes $p,q,r,s$ are used otherwise.

It is useful in orbital dependent approaches to 
divide the XC energy as $E_{\mathrm{xc}}=E_\mathrm{x}+E_\mathrm{c}$, separating the 
exchange and the correlation contributions.
The exchange energy functional has the form
of the usual Hartree--Fock exchange energy
\begin{equation}
E_\mathrm{x}[\{\phi_{q\tau}\}] = -\frac{1}{2}\sum_\sigma\sum_{ij}(i_\sigma j_\sigma|j_\sigma i_\sigma)\ ,
\end{equation}
with $(p_\sigma q_\sigma|r_\sigma s_\sigma)$ being a two--electron integral in the Mulliken 
notation computed from KS orbitals.
The corresponding OEP KS exact--exchange potential is labeled OEPx and defined by the
equation
\begin{eqnarray} \label{vx}
v_{\mathrm{x},\sigma}^{\mathrm{OEPx}}(\R) &= -2\sum_{ij}\sum_a \int \frac{(i_\sigma j_\sigma|j_\sigma a_\sigma)}{\epsilon_{i\sigma}-\epsilon_{a\sigma}} \nonumber\\
& \times \phi_{a\sigma}(\R')\phi_{i\sigma}(\R')X^{-1}_\sigma(\R',\R)d\R'\ .
\end{eqnarray}

For the correlation part, we consider the spin--resolved 
expression obtained from the second--order G\"orling--Levy perturbation theory (GLPT)
energy functional \cite{gorling:1993:PT}, which has exactly the same form as a functional
defined from the many--body perturbation theory (MBPT) \cite{grabowski:2002:OEPP2,ivanov:2003:P2}
\begin{equation}\label{e6}
E_c^{(2)} = c_{\mathrm{OS}}E_{\mathrm{OS}} +  c_{\mathrm{SS}}E_{\mathrm{SS}}  + E_{\mathrm{SE}}\ ,
\end{equation}
where
\begin{eqnarray}
E_{\mathrm{OS}} & = & \frac{1}{2}\sum_\sigma\sum_{\tau\ne\sigma}\sum_{ij}\sum_{ab}\frac{\left|(i_\sigma a_\sigma|j_\tau b_\tau)\right|^2}{\epsilon_{i\sigma}+\epsilon_{j\tau}-\epsilon_{a\sigma}-\epsilon_{b\tau}}
\label{eOS}\\
E_{\mathrm{SS}} & = & \frac{1}{2}\sum_\sigma\sum_{ij}\sum_{ab}\frac{\left|(i_\sigma a_\sigma|j_\sigma b_\sigma)\right|^2}{\epsilon_{i\sigma}+\epsilon_{j\sigma}-\epsilon_{a\sigma}-\epsilon_{b\sigma}}
 \nonumber\\
&& -\frac{1}{2}\sum_\sigma\sum_{ij}\sum_{ab}\frac{(i_\sigma a_\sigma|j_\sigma b_\sigma)(a_\sigma j_\sigma|b_\sigma i_\sigma)}{\epsilon_{i\sigma}+\epsilon_{j\sigma}-\epsilon_{a\sigma}-\epsilon_{b\sigma}}
\label{eSS}\\
E_{\mathrm{SE}} & = & \sum_\sigma\sum_{ia}\frac{|f^\sigma_{ia}|^2}{\epsilon_{i\sigma}-\epsilon_{a\sigma}} ,
\label{es}
\end{eqnarray}
while $c_{\mathrm{OS}}$ and $c_{\mathrm{SS}}$ are simple scaling factors for 
the opposite--spin (OS) and same--spin (SS)
correlation, respectively. Note that the correlation
functional of Eq. (\ref{e6}) includes also a 
 singly excited
 term,
depending on the square of the Fock matrix elements 
\begin{equation}
f^\sigma_{pq}=\varepsilon_{p_\sigma}^{\mathrm{KS}}\delta _{p_\sigma
q_\sigma}-\langle
p_\sigma|\hat{K}_\sigma+v_{\mathrm{x},\sigma}^{OEP}|q_\sigma\rangle,
\end{equation}
with $\hat{K}$ the nonlocal Hartree--Fock exchange operator. This term
arises because in the present case the correlation energy through second--order  
is computed using KS orbitals instead of canonical Hartree--Fock ones.
The OEP correlation potential corresponding to the functional
of Eq. (\ref{e6}) is obtained from Eq. (\ref{e1}) as
\begin{equation}\label{e11}
v_{\mathrm{c},\sigma}^{\mathrm{OEP}}(\R) = 
 c_{\mathrm{OS}}v_{\mathrm{OS},\sigma}(\R) +
c_{\mathrm{SS}}v_{\mathrm{SS},\sigma}(\R)
+ v_{\mathrm{SE},\sigma}(\R)\ ,
\end{equation}
with
\begin{eqnarray}
\nonumber
v_{\mathrm{OS},\sigma}(\R) & = & \sum_{\tau\ne\sigma}\sum_{ij}\sum_{ab}\int\Bigg\{\frac{(i_\tau a_\tau|j_\sigma b_\sigma)}{\epsilon_{i\tau}+\epsilon_{j\sigma}-\epsilon_{a\tau}-\epsilon_{b\sigma}} \\
\nonumber
&& \times\bigg[2\sum_{p\ne j}\frac{(i_\tau a_\tau|p_\sigma b_\sigma)}{\epsilon_{j\sigma}-\epsilon_{p\sigma}}\phi_{j\sigma}(\R')\phi_{p\sigma}(\R')  \\
\nonumber
&&+ 2\sum_{p\ne b}\frac{(i_\tau a_\tau|j_\sigma p_\sigma)}{\epsilon_{b\sigma}-\epsilon_{p\sigma}}\phi_{p\sigma}(\R')\phi_{b\sigma}(\R')  \\
\nonumber
&&-\frac{(i_\tau a_\tau|j_\sigma b_\sigma)}{\epsilon_{i\tau}+\epsilon_{j\sigma}-\epsilon_{a\tau}-\epsilon_{b\sigma}}\Big(\phi_{j\sigma}(\R')\phi_{j\sigma}(\R')\\
&& - \phi_{b\sigma}(\R')\phi_{b\sigma}(\R')\Big)\bigg]\Bigg\}X_\sigma^{-1}(\R',\R)d\R'
\end{eqnarray}
\begin{eqnarray}
\nonumber
v_{\mathrm{SS},\sigma}(\R) & = & \sum_{ij}\sum_{ab}\int\Bigg\{ \frac{(i_\sigma a_\sigma|j_\sigma b_\sigma)-(a_\sigma j_\sigma|b_\sigma i_\sigma)}{\epsilon_{i\sigma}+\epsilon_{j\sigma}-\epsilon_{a\sigma}-\epsilon_{b\sigma}} \\
\nonumber
&& \times \bigg[2\sum_{p\ne i}\frac{(p_\sigma a_\sigma|j_\sigma b_\sigma)}{\epsilon_{i\sigma}-\epsilon_{p\sigma}}\phi_{i\sigma}(\R')\phi_{p\sigma}(\R')  \\
\nonumber
&&+ 2\sum_{p\ne a}\frac{(i_\sigma p_\sigma|j_\sigma b_\sigma)}{\epsilon_{a\sigma}-\epsilon_{p\sigma}}\phi_{p\sigma}(\R')\phi_{a\sigma}(\R')  \\
\nonumber
&&-\frac{1}{2}\frac{(i_\sigma a_\sigma|j_\sigma b_\sigma)}{\epsilon_{i\sigma}+\epsilon_{j\sigma}-\epsilon_{a\sigma}-\epsilon_{b\sigma}}\Big(\phi_{i\sigma}(\R')\phi_{i\sigma}(\R')\\
\nonumber
&& + \phi_{j\sigma}(\R')\phi_{j\sigma}(\R') - \phi_{a\sigma}(\R')\phi_{a\sigma}(\R')  \\
&& - \phi_{b\sigma}(\R')\phi_{b\sigma}(\R') \Big)\bigg]\Bigg\}X_\sigma^{-1}(\R',\R)d\R' 
\end{eqnarray}
\begin{eqnarray}
\nonumber
v_{\mathrm{SE},\sigma}(\R) & = & 2\sum_{ia}\int\Bigg\{\frac{f^\sigma_{ia}}{\epsilon_{i\sigma}-\epsilon_{a\sigma}}\bigg[\sum_{p\ne i}\frac{\phi_{p\sigma}(\R')\phi_{i\sigma}(\R')}{\epsilon_{i\sigma}-\epsilon_{p\sigma}}f^\sigma_{pa}  \\
\nonumber
&& + \sum_{p\ne a}\frac{\phi_{p\sigma}(\R')\phi_{a\sigma}(\R')}{\epsilon_{a\sigma}-\epsilon_{p\sigma}}f^\sigma_{ip} + \sum_{kc}\frac{\phi_{c\sigma}(\R')\phi_{k\sigma}(\R')}{\epsilon_{k\sigma}-\epsilon_{c\sigma}}\\
\nonumber
&& \times \Big(2(i_\sigma a_\sigma|c_\sigma k_\sigma) - (i_\sigma c_\sigma|k_\sigma a_\sigma) - (i_\sigma k_\sigma|c_\sigma a_\sigma)\Big) \\
\nonumber
&& - \frac{1}{2}\frac{f^\sigma_{ia}}{\epsilon_{i\sigma}-\epsilon_{a\sigma}}\Big(\phi_{i\sigma}(\R')\phi_{i\sigma}(\R')-\phi_{a\sigma}(\R')\phi_{a\sigma}(\R')\Big)\bigg]\Bigg\} \\
&& \times X^{-1}_\sigma(\R',\R)d\R'\ . \label{ves}
\end{eqnarray}

\section{Computational methodology}\label{sec:Comp}
 Equations (\ref{vx}) and (\ref{e11})-(\ref{ves}) provide formal
expressions for the exchange and correlation potentials, respectively.
However,
to calculate the OEP exchange and correlation potentials in practice we employ the
finite--basis set
procedure of Refs. \onlinecite{gorling:1999:OEP,ivanov:1999:OEP,ivanov:2002:OEP}.
Thus, the OEP is expanded as
\begin{equation}\label{ee18}
v_{\mathrm{xc},\sigma}^{\mathrm{OEP}}(\R) = v_{\mathrm{Slater}}^\sigma(\R) + \sum_{l=1}^{N_{\mathrm{aux}}}c_l^\sigma g_l(\R)\ ,
\end{equation}
 and this ansatz is used to solve the OEP equation (Eq. (\ref{e1})).
The first term on the right hand side is the Slater potential
\begin{equation}
v_{\mathrm{Slater}}^\sigma(\R) = - \sum_{ij}\frac{\phi_{i\sigma}(\R)\phi_{j\sigma}(\R)}{\rho_\sigma(\R)}\int\frac{\phi_{i\sigma}(\R')\phi_{j\sigma}(\R')}{|\R-\R'|}d\R' 
\end{equation}
which is added to ensure the correct $-1/r$ asymptotic behaviour of the potential.
The second term is an expansion in $N_{\mathrm{aux}}$ Gaussian basis functions $g_l(\R)$ with
the expansion coefficients $c_l^\sigma$ determined from the solution of the OEP 
equations 
  with the total
$v_{\mathrm{xc},\sigma}^{\mathrm{OEP}}=v_{\mathrm{x},\sigma}^{\mathrm{OEP}}+v_{\mathrm{c},\sigma}^{\mathrm{OEP}}$
potentials defined by Eq. (\ref{vx}) and Eqs. (\ref{e11})-(\ref{ves}) for exchange and
correlation potentials respectively.

Note that Eq. (\ref{ee18})  cannot correctly describe the exact OEP potential in
 the asymptotic region for molecules
with HOMO nodal surfaces (HNSs), i.e.  H$_2$O, NH$_3$, C$_2$H$_6$ in our
 test set. In fact, for these molecules
the exact OEP potential is characterized by asymptotic barrier-well
 structures near HNSs \cite{lhfasy,prlasy}  which cannot be
represented in a (finite) linear combination of Gaussian functions.
As shown in Refs. \onlinecite{lhfasy,prlasy}, the asymptotic
 barrier-well structures will
 significantly influence high-lying virtual orbitals (LUMO+1 and above) so
that
the total correlation energies and correlation potentials can be expected
 to be influenced as well. We computed the GL2 correlation energy for
C$_2$H$_6$
using localized Hartrre-Fock (LHF) \cite{lhf1} orbitals with and without the correct treatment of
 the asymptotic region \cite{lhfasy,prlasy} and we found a very small difference (about 1
mH).
However, the description of the asymptotic region of correlated OEP method
is beyond the target on this work.

 Numerical instabilities in the solution of the OEP equations
\cite{hirata:2001:OEPU,elp,hesselmanexx07,Joubert1,Kollmar1,Teale1,Glushkov1,Theophilou1,grabowski:2005:1shot,grabowski:2007:ccpt2,bulat:2007:sn}
were minimized by a careful choice of the basis set
(see Sec. \ref{comp_det} for further details). Our calculations employ a truncated
 singular-value decomposition (SVD) for the construction of the pseudo-inverse of
 the linear response function in the OEP procedure. This regularization is
 an essential step in determining stable solutions to Eq. (\ref{e1}) and, together
 with the choice of basis set, ensures that stable and physically sound
 solutions are obtained.

In the present work we consider a full
family of spin--component--scaled (SCS) second-order 
OEP (OEP2) methods, obtained using the correlation energy 
contributions in Eqs.~(\ref{e6})--(\ref{es}) and the 
corresponding correlation potentials of 
Eqs. (\ref{e11})--(\ref{ves}) with different values 
of the parameters $c_{\mathrm{OS}}$ and $c_{\mathrm{SS}}$ .
These methods are denoted in general as OEP2--SCS.
Moreover, special attention is devoted to OEP approaches based only 
on the opposite--spin part of the correlation, which are generally
labeled as OEP2--SOS and described in more detail in 
Sections \ref{sec:ResOS1} and \ref{sec:ResOS2}.
Finally, we consider, for comparison also the OEP--GL2 
\cite{grabowski:2002:OEPP2,engel:2005:oeppt2,mori-sanchez:2005:oeppt2}
and OEP2--sc \cite{bartlett:2005:abinit2} methods.
We note that the former is also a member of the
family of the OEP2--SCS methods, being obtained by
setting $c_{\mathrm{SS}}=c_{\mathrm{OS}}=1$.

Finally, we point out that, unfortunately, our current pilot
OEP2-SOS implementation in ACESII \cite{acesii} does
not have the  $\mathcal{O}(N^4)$  scaling yet enabled.
This does not
significantly impact the efficiency of the calculations
presented here for relatively small systems, however, it is an important
advantage for larger systems and basis sets.

\subsection{Computational details}

\label{comp_det}
To test the various OEP2--SCS approaches we performed calculations on 
several atomic (He, Be, Ne, Mg, Ar) and molecular 
(H$_2$, He$_2$, HF, CO, H$_2$O, Cl$_2$, N$_2$, Ne$_2$ HCl,  
NH$_3$, C$_2$H$_6$) systems 
using the the ACES II package \cite{acesii}.
For all systems we considered equilibrium geometries from
Refs.~\onlinecite{herzberg70,boucier70,callonion76}. 
The geometries are as follows: H$_2$ H--H = 0.7461\AA; He$_2$ He--He =
5.6\AA; N$_2$ N--N = 1.098\AA; Ne$_2$ Ne--Ne = 3.1\AA; HF H--F =
0.9169\AA; HCl H--Cl = 1.2746\AA; Cl$_2$ Cl--Cl = 1.9871\AA; CO C--O =
1.128\AA; H$_2$O H--O = 0.959\AA, H--O--H = 103.9$^{\circ}$ ; NH$_3$
N--H = 1.008\AA, H--N--H = 111.552$^{\circ}$  ; C$_2$H$_6$ C--C =
1.533\AA, C--H = 1.107\AA, H--C--C = 109.3$^{\circ}$ , H-C-H =
109.642$^{\circ}$ .

In the present work we use the same basis set for the representation of both
the orbitals and the OEP potential (i.e. the principal and the auxiliary
 basis sets are the same). Whilst this combination may not be optimal in
 terms of the balance between the potential and orbital descriptions
\cite{hesselmanexx07}
 it has been shown to give reasonable
 results and a rationale for this has been presented by Filatov et al. in 
Ref.~\onlinecite{filatov10}. To ensure that the basis sets chosen were flexible enough for both
 representation of the orbitals and potentials, all basis sets were
 constructed by  full uncontraction
of medium size (triple
zeta) basis sets originally developed for correlated calculations as in 
Refs. \onlinecite{grabowski:2011:jcp,grabowski:2013:molphys}. In particular,
we employed an even tempered $20s10p2d$ basis for He atom and He$_2$ molecule, 
an uncontracted ROOS--ATZP basis~\cite{widm90} for Be,  
Ne atom and Ne$_2$ molecule, and an uncontracted aug--cc--pVTZ basis set~\cite{woon93} for Mg atom.
For the Ar atom we used a modified basis set which combines $s$ and $p$ type basis functions 
from the uncontracted ROOS--ATZP~\cite{widm90} with $d$ and $f$ functions coming from the uncontracted 
aug--cc--pwCVQZ basis set~\cite{peterson02}. 
The uncontracted cc--pVTZ basis set of Dunning~\cite{dunning:1989:bas} was used for all other systems.

For all OEP--KS calculations tight convergence criteria were enforced, e.g. for the SCF, 
corresponding to maximum deviations in density
matrix elements of 10$^{-8}$ au.
 In addition the truncated SVD
cutoff was set to 10$^{-6}$
and results were carefully
checked to ensure convergence with respect to this parameter. As a further
test of the stability of our results we computed the  gradient of the total
electronic energy with respect to variations of the $c_{l}^\sigma$  coefficients in Eq.
(\ref{ee18}): in all cases the computed gradient had a norm less than
10$^{-12}$ and the energies computed at the perturbed coefficients
were higher than those obtained in our converged calculations, consistent
with an energy minima.

To assess the OEP results we considered reference data from the OEP2--sc
method \cite{bartlett:2005:abinit2} and benchmark data from 
second--order M\o ller--Plesset perturbation theory (MP2) \cite{moll34},
scaled--opposite--spin (SOS) MP2 \cite{jung04},
and coupled cluster singles--doubles with perturbative triples [CCSD(T)] 
\cite{:/content/aip/journal/jcp/76/4/10.1063/1.443164,:/content/aip/journal/jcp/89/12/10.1063/1.455269,:/content/aip/journal/jcp/87/10/10.1063/1.453520,Raghavachari1989479}
calculations. In the MP2 and CCSD(T) cases relaxed densities were obtained from relaxed 
density matrices 
\cite{hansch:1984:zvec,ricamo:1985:relden,bartlett:1986:relden,bartlett:1989:relden}
constructed using the Lagrangian approach 
\cite{helg:1989:lag,jorg:1988:lag,koch:1990:lag,hald:2003:lag}, 
while KS potentials and the corresponding
single--particle orbitals were constructed by employing 
the inversion approach of Wu and Yang \cite{wu:2003:wy}. In these calculations the
smoothing--norm approach of 
Refs.~\onlinecite{heaton:2007:sn,bulat:2007:sn} was employed with a regularization parameter of $10^{-5}$. The calculations were considered converged when the gradient norm was below $10^{-6}$.
The results of this inversion approach when applied to MP2 and CCSD(T) relaxed densities 
were denoted by KS[MP2] and KS[CCSD(T)] respectively.
All benchmark calculations were carried out with a development version of the
\textsc{Dalton2013} quantum chemistry program \cite{DALTON,DALPAP}.

To assess the performance of each approach we considered 
the following quantities relative to CCSD(T) and KS[CCSD(T)].

i) Absolute differences in the correlation energy
\begin{eqnarray}
\Delta E_\mathrm{c} & = & \left|E_\mathrm{c}^{\mathrm{method}}-E_\mathrm{c}^{\mathrm{CCSD(T)}}\right|
\end{eqnarray}
Note that we compared DFT correlation energies with WFT ones; 
see discussion in Section \ref{sec:ResOS2}.

ii) Absolute differences in the energy gap between the lowest
unoccupied molecular orbital (LUMO) and the highest occupied molecular
orbital (HOMO)
\begin{eqnarray}
\Delta \mathrm{HL} & = & \Big|\epsilon_{\mathrm{L}}^{\mathrm{method}} - \epsilon_\mathrm{H}^{\mathrm{method}}  \\
\nonumber
&& -\left(\epsilon_{\mathrm{L}}^{\mathrm{KS[CCSD(T)]}} - \epsilon_{\mathrm{H}}^{\mathrm{KS[CCSD(T)]}}\right)\Big|\ .
\end{eqnarray} 
 We remark that this quantity is an important indicator
because it is not only directly related to the quality of the
OEP potential, which determines the orbital energies, but also
because within time-dependent DFT the HOMO-LUMO gap is
the zero-th order approximation for excitation energies.
We also note that all of our calculations deliver LUMO orbitals 
that are bound,
 this is consistent with the fact that the KS equations (in contrast to the
 Hartree--Fock ones) contain the same local self-interaction free potential
 for all orbitals (both occupied and virtual).

iii) The integrated density differences (IDD)
\begin{eqnarray}
\mathrm{IDD} & = & \int_{r_0}^{+\infty} \Omega(r)\left|\rho^{\mathrm{method}}(r) - \rho^{\mathrm{CCSD(T)}}(r)\right| dr
\label{eidd}
\end{eqnarray}
where $\Omega(r)=4\pi r^2$ and $r_0=0$ for atoms, while
for linear molecules $\Omega(r)=1$, $r_0=-\infty$ and $r$ is the distance 
along the main axis .
We note that the IDD is directly related in atoms and molecules
with  the difference radial
 density distribution \cite{Jankowski:2009:DRD}
$\mathrm{DRD}(r)=4\pi r^2(\rho^{\mathrm{method}}(r)-\rho^{\mathrm{xref}}(r))$
and the difference total density distribution (correlated density)
$\rho_c(r)=(\rho^{\mathrm{method}}(r)-\rho^{\mathrm{xref}}(r))$,
respectively, where xref=OEPx for correlated OEP methods and xref=Hartree--Fock
for conventional wave function correlated methods. 
Thus, it provides also a direct test of the quality of the
correlation potential. 

\section{Results}\label{sec:Res}
In this section we present the results of various OEP2-SCS 
calculations and compare them with benchmark values. 
In particular, we analyse the importance of 
 the singly excited term
for both the energies and the potentials and 
the role of the different values of the 
scaling parameters  $c_{\mathrm{OS}}$ and $c_{\mathrm{SS}}$.

\subsection{Role of the singly excited term}\label{sec:ResSE}
To inspect the role of the 
 singly excited
term in the
second--order correlation energy we report in Table \ref{tab1}
the decomposition of the energy according to Eq. (\ref{e6}) 
for several systems.
\begin{table}
\caption{\label{tab1} Total correlation energy (m$E_\mathrm{h}$) and its components according to Eq. (\ref{e6}) for
OEP--GL2 calculations on several atoms and molecules. 
In the last line we report the percentage contribution of each term to the total correlation energy.}
\begin{ruledtabular}
\begin{tabular}{lrrrr}
System & $E_c^{\mathrm{OEP-GL2}}$ & $E_{\mathrm{SS}}^{\mathrm{OEP-GL2}}$ & $E_{\mathrm{OS}}^{\mathrm{OEP-GL2}}$ &$E_{\mathrm{SE}}^{\mathrm{OEP-GL2}}$ \\
\hline
He     &  -46.24 &    0.00 &  -46.24 &  0.00 \\ 
Ne     & -457.82 & -113.43 & -342.77 & -1.61 \\ 
Ar     & -776.35 & -205.02 & -565.86 & -5.46 \\ 
Mg     & -277.20 &  -54.56 & -219.35 & -3.28 \\ 
H$_2$  &  -48.55 &    0.00 &  -48.55 &  0.00 \\ 
He$_2$ &  -92.47 &    0.00 &  -92.47 &  0.00 \\ 
HF     & -464.76 & -113.35 & -349.13 & -2.28 \\ 
CO     & -863.12 & -210.34 & -646.91 & -5.87 \\ 
H$_2$O & -466.33 & -108.76 & -354.95 & -2.63 \\ 
HCl    & -516.48 & -125.58 & -385.10 & -5.80 \\ 
Cl$_2$ &-1025.12 & -255.99 & -753.73 &-15.40 \\ 
N$_2$  & -866.33 & -210.47 & -649.93 & -5.93 \\ 
Ne$_2$ & -915.95 & -226.99 & -685.71 & -3.26 \\ 
NH$_3$ & -421.59 & -90.13  & -329.01 & -2.45 \\ 
C$_2$H$_6$ & -693.66 & -134.40 & -552.81 & -6.44 \\ 
       &         &         &         & \\
\%     &         & 18.8\%  & 80.6\%  & 0.6\% \\
\end{tabular}
\end{ruledtabular}
\end{table}
Inspection of the table shows that the contribution of the
 singly excited
term is indeed very small, amounting to a
few m$E_{\mathrm{h}}$ in most systems, with the notable exception of Cl$_2$
where $E_{\mathrm{SE}}^{\mathrm{OEP-GL2}}$=$-$15.40 m$E_{\mathrm{h}}$ (
anyway
 this is only 1.5\%
of the total $E_c^{\mathrm{OEP-GL2}}$ energy). Thus, the  singly excited
term contributes only slightly to the total OEP correlation energy,
giving on average a contribution lower than 1\%. More importantly,
this contribution is much smaller than the typical error
of the OEP--GL2 method with respect to accurate benchmark correlation energies (e.g.
CCSD(T)). For most systems, in fact, the errors relative to CCSD(T) are found to be 
on the order of several tens of m$E_{\mathrm{h}}$ (see Table \ref{tab2}) and for Cl$_2$ the error in 
the $E_c^{\mathrm{OEP-GL2}}$ energy is more than 150 m$E_{\mathrm{h}}$. 

\begin{figure}
\begin{center}
\includegraphics[width=\columnwidth,angle=270]{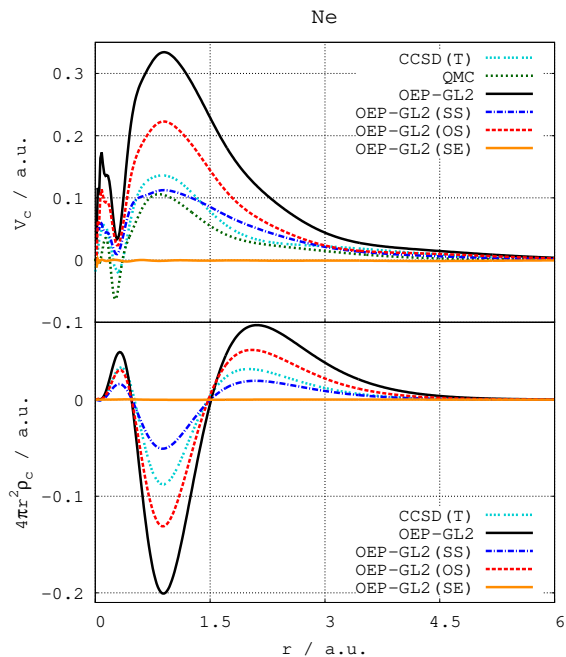}
\caption{\label{fig1} Correlation potentials (top) and 
DRD (=$4\pi\rho_{\mathrm{c}}$)
(bottom) for the Ne atom 
as resulting from different components of the
$v_{\mathrm{c}}^{\mathrm{OEP-GL2}}$ potential. }
\end{center}
\end{figure}

An even less important role is played by the 
 singly excited
term
when the correlation potential is considered. This is
shown in Figure~\ref{fig1} where we report the
plot of the correlation KS potential and the DRD for the Neon atom, as a typical case.
Similar results are obtained
for  other systems.
The figure shows that, on the scale of the plots, the contribution of the
 singly excited term
to the correlation potential is almost negligible
Thus, almost no effect can be expected from the
 singly excited term on density--related and single--particle properties
(e.g. multipole moments or orbital energies).

This result apparently contrasts the usual finding for the calculation of
MP2 relaxed densities, where the single excitations display a non-negligible
role (note that such contributions are included in all MP2 density calculations in this work). 
This difference can be rationalized considering that in the case of
MP2 (based on fixed Hartree--Fock orbitals) the single excitations must
account for significant orbital relaxation effects. In contrast, in the
present OEP2- calculations the orbitals are already optimized to minimize
the second--order energy expression, hence a large part of the contribution
from the single excitations is no longer necessary. The remaining small contribution from the singly excited term
in the OEP2- methods is a measure of the difference between the KS and the
Hartree--Fock orbitals. The present results show that the 
singly excited term gives a contribution to the correlation energy that is at least two orders of magnitude
smaller than that of the doubly excited term (OS+SS). The results of this analysis suggest that in most cases
the singly excited term can be safely neglected in correlated OEP2--SCS
calculations, and so we shall neglect them throughout the remainder of this work.

\subsection{Two--dimensional scan of the $c_{\mathrm{OS}}$
and $c_{\mathrm{SS}}$ parameters}\label{sec:ResDoub}

In the previous subsection we showed that the major contribution
to the OEP second--order correlation comes from the
doubly--excited terms of Eqs. (\ref{e6}) and (\ref{e11}), that describe
the same--spin and opposite--spin correlation contributions.
Several studies have suggested that these two contributions are
not completely independent, but rather 
approximately proportional to each other, and that the
overall description of the correlation might be improved by a proper scaling
of the two terms \cite{sos_oep2:prb,C3CP51431E,WCMS12}.
In fact, as shown in Figure~\ref{fig1}, the SS and OS potential and DRD, show a very high degree of proportionality,
and a proper scaling of the SS potential can reproduce the OS potential\cite{sos_oep2:prb}.
However, so far, this proportionality has been been investigated 
only in a qualitative manner, we now perform a more
quantitative analysis. We have performed a full two--dimensional scan of the space spanned
by the $c_{\mathrm{OS}}$ and $c_{\mathrm{SS}}$ parameters, analysing the resulting errors
on the indicators introduced in Section \ref{comp_det}.
\par
\begin{figure*}
\begin{center}
\includegraphics[width=0.9\textwidth,angle=270]{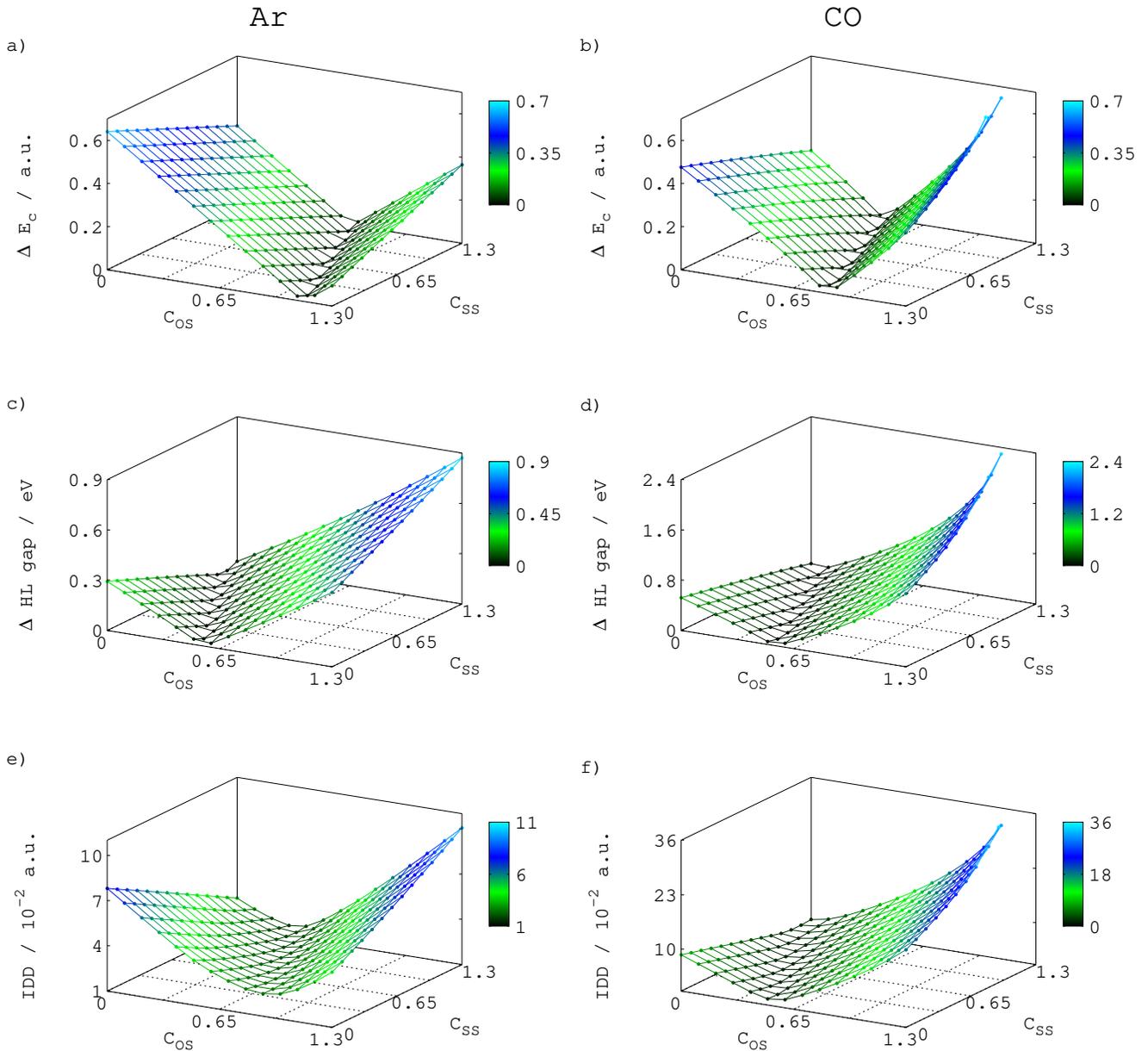}
\caption{\label{fig2} Scans of $\Delta E_\mathrm{c}$, $\Delta \mathrm{HL}$ and $\mathrm{IDD}$ with respect
 to the values of the $c_{\mathrm{OS}}$ and $c_{\mathrm{SS}}$ parameters for the Ar atom (left column) 
and the CO molecule (right column).
}
\end{center}
\end{figure*}

The outcome of this study is summarized in Figure~\ref{fig2}
where we report, as representative examples, the results for 
the Ar atom and the CO molecule.
Similar results (not reported) have been obtained for Ne, Mg, HF, and N$_2$. 
Note that each point in the  ($c_{\mathrm{OS}}$, $c_{\mathrm{SS}}$) space corresponds 
to a full SCF OEP--SCS calculation.

The plots show three main features: 
\begin{enumerate}

\item For $\Delta E_\mathrm{c}$ and  $\Delta \mathrm{HL}$ there is no unique
pair of  $c_{\mathrm{OS}}$ and $c_{\mathrm{SS}}$ values that minimizes the error; on the contrary, it is
possible to identify a continuous set of parameter pairs, defined by a linear
relation of the type $c_{\mathrm{OS}} = ac_{\mathrm{SS}} + b$, with $a$ and $b$
system-- and
property--dependent constants, for which the error in minimized.
The values on the minimizing line are exactly zero: in fact the plots show the absolute error 
and the signed error is positive (negative) for small (large) values of $c_{\mathrm{OS}}$.
The fact that the error is exactly zero is not surprising because $\Delta E_\mathrm{c}$ and $\Delta \mathrm{HL}$ are  
{\it single--valued} quantities, and thus, for a given system and a given property, it is always 
possible to scale the  $c_{\mathrm{OS}}$ and $c_{\mathrm{SS}}$ so that the exact reference values are reproduced.

\item For the $\mathrm{IDD}$, the situation is different. In fact, the $\mathrm{IDD}$ is an  integrated quantity 
which describes how correlation effects
are reproduced in the density for {\it all} points 
in
space. Thus the error in general is non-zero, 
because the CCSD(T) reference correlated density
is not easily reproduced by a correlation potential which 
includes
 only
second--order contributions.
The plots in the bottom panels of Figure~\ref{fig2} show in fact  that the absolute minima can be 
obtained for   
$c_{\mathrm{OS}}=0.2$,  
$c_{\mathrm{SS}}=1.5$ for Argon  and 
$c_{\mathrm{OS}}=0.4$, 
$c_{\mathrm{SS}}=0.3$ for CO.
However, the plots still have a parabolic profile, indicating that for each fixed choice of $c_{\mathrm{SS}}$ a single value of $c_{\mathrm{OS}}$ can be chosen to minimize the $\mathrm{IDD}$ within this constraint.
  
\item Setting $c_{\mathrm{SS}}=0$ it is possible to find a value of the $c_{\mathrm{OS}}$ parameter
such that the error for any given property 
is close to the absolute minimum. Thus, it is meaningful to consider
a scaled--opposite--spin (SOS) second--order OEP method and expect that it may
exhibit a performance close to the best achievable by second--order
correlation, such approaches can be computationally more favourable and are discussed in the next section.

\end{enumerate}

\subsection{Role of the $c_{\mathrm{OS}}$ parameter in OEP2--SOS methods}\label{sec:ResOS1}
In this section we focus our attention on scaled--opposite--spin
second--order OEP approaches, which are the most appealing for
practical computational applications.
We recall in fact that SOS methods provide a computational advantage
with respect to other second--order correlation approaches, because
when properly implemented they scale as $\mathcal{O}(N^4)$ rather than as
$\mathcal{O}(N^5)$.
 In this
case, the computationally expensive calculation of the
exchange integrals appearing in the SS part is avoided and
an efficient  $\mathcal{O}(N^4)$  scaling of the SOS-OEP2 method can be obtained.\cite{jung04}

In particular, we investigate the performance of the OEP2--SOS
correlation methods as a function of the parameter
$c_{\mathrm{OS}}$.

\begin{figure*}
\begin{center}
\includegraphics[width=0.5\textwidth,angle=270]{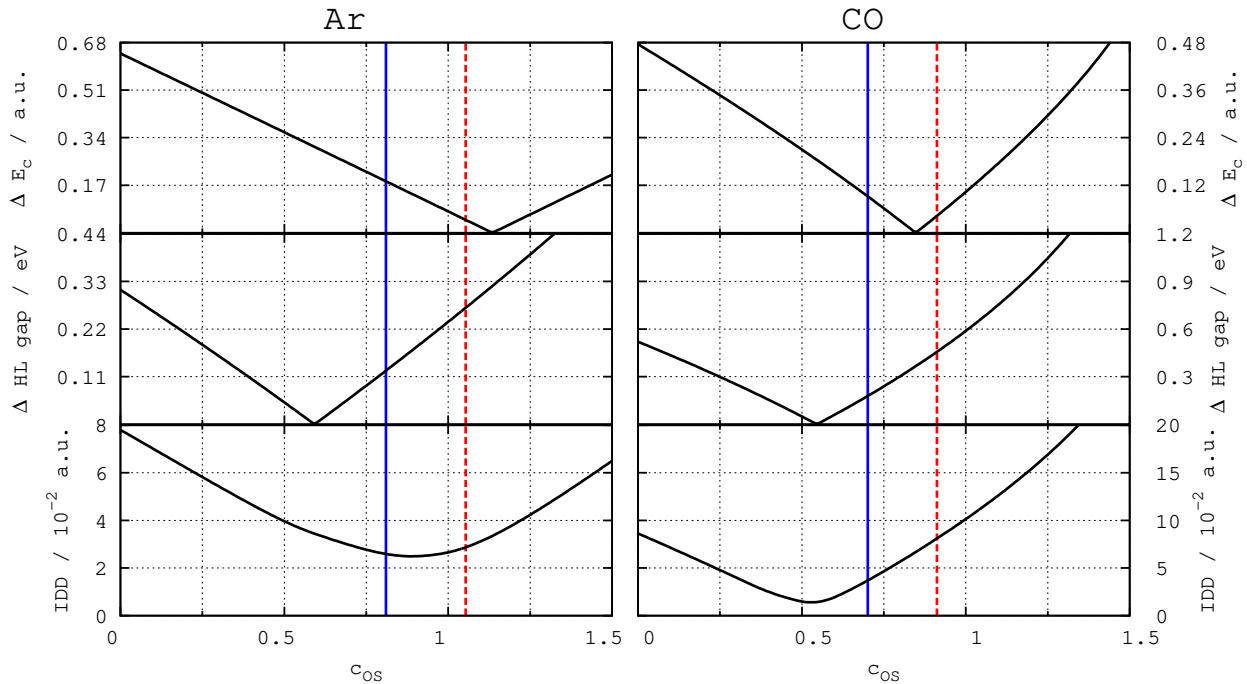}
\caption{\label{fig3} Scans of the errors in several properties (see Section \ref{comp_det})
 as obtained from OEP2--SOS approaches with different values of the
 $c_{\mathrm{OS}}$ parameter, for Ar atom (left panel) and CO (right panel). 
 The red (dashed) and blue vertical lines show the values of $c_{\mathrm{OS}}$ parameter calculated for the 
OEP2--SOSa and OEP2--SOSb approaches (see Eqs. (\ref{cOS1}) and (\ref{cOS2})), respectively.}

\end{center}
\end{figure*}
Figure~\ref{fig3} reports, for two typical cases, 
the errors in different properties (see Section
\ref{comp_det} for the definitions of the indicators) 
as obtained from the OEP2--SOS approach using different values of the $c_{\mathrm{OS}}$
parameter.
The figure shows that in all cases a smooth curve is obtained and a single
minimum can be well identified for each property. 
Nevertheless, the minima occur at different
values of the $c_{\mathrm{OS}}$ parameter for different systems and properties.
Thus, it is not possible to determine with accuracy a single ``best'' value
for the opposite--spin scaling parameter $c_{\mathrm{OS}}$.

As suggested in Ref. \onlinecite{sos_oep2:prb}, a reasonable description of the
correlation energy is obtained in most cases by using 
a system--dependent value of $c_{\mathrm{OS}}$, 
\begin{equation}
\label{cOS1}
c_{\mathrm{OS}}^{\mathrm{a}} = \frac{1.3\ E_{\mathrm{OS}}@\mathrm{HF}}{E_{\mathrm{OS}}@\mathrm{OEPx}}\ ,
\end{equation}
where $E_{\mathrm{OS}}@\mathrm{HF}$ and $E_{\mathrm{OS}}@\mathrm{OEPx}$ denote the opposite 
spin part (see Eq.(\ref{eOS})) of the second--order correlation
energy expression Eq. (\ref{e6}) computed with Hartree--Fock and OEP
exchange--only
orbitals, respectively. Using this value of the
scaling parameter we can thus define the OEP2-SOSa method 
(this approach was denoted OEP-SOS in  Ref. \onlinecite{sos_oep2:prb})   
designed to yield accurate correlation energies. 
 Thus, in the numerator of
Eq. (\ref{cOS1}) we
aim  to reproduce the reference 
(CCSD(T)) correlation energy with a noniterative calculation and according
to Ref. \onlinecite{jung04}
this can be achieved (approximately) by setting 
$E_{REF} \approx 1.3 E_{OS}@HF$.

On the other hand, 
analysing the behaviour of different properties as functions of
the $c_{OS}$ parameter in several systems,
we found that the $c_{\mathrm{OS}}^{\mathrm{a}}$ values are in general
too large to achieve an accurate description of the HOMO-LUMO gap and correlated densities. 
In fact, we found  that
these are better 
described using the system-dependent parameter
\begin{equation}
\label{cOS2}
c_{\mathrm{OS}}^{\mathrm{b}} = \frac{E_{\mathrm{OS}}@\mathrm{HF}}{E_{\mathrm{OS}}@\mathrm{OEPx}}\ .
\end{equation}
This value of the scaling parameter can thus be used to define
the OEP2-SOSb method, which is, by construction, optimized
for the description of correlation-potential-related properties.
 
We note that, there is in practice
no additional numerical cost necessary
for calculating the value of the  $c_{\mathrm{OS}}$ parameter.
In fact, as pointed out in Ref. \onlinecite{sos_oep2:prb},
Hartree--Fock and OEPx calculations can be considered as
intermediate steps in any OEP2 calculation.
 We note also that, despite being system dependent,
our $c_{OS}$ coefficients define OEP2-SOS methods which are 
 size extensive, because the coefficients 
are constructed as ratios of two size-extensive quantities (Eqs. (\ref{cOS1}) 
and (\ref{cOS2})).

\subsection{Assessment of different OEP2-SOS methods}\label{sec:ResOS2}
In the previous subsection we saw that it is possible
to define two OEP2-SOS approaches, which are expected to work
rather accurately for a set of properties.
In this section we now  assess these methods in a systematic
manner to understand their merits and limitations.
Moreover, we consider a third approach, namely the
OEP2-SOSh method, which is a hybrid of those in Section~\ref{sec:ResOS1}.
Within the OEP2-SOSh method the self-consistent-field
calculations are performed using the $c_{\mathrm{OS}}^{\mathrm{b}}$ coefficient,
so that accurate potentials, densities and other one--electron quantities 
 can be obtained, but the
final correlation energies are scaled by a factor 1.3, so that
they roughly correspond to those of the OEP2-SOSa method obtained with the
$c_{\mathrm{OS}}^{\mathrm{a}}$ coefficient.

To start our assessment we report in Table \ref{tab2}
the correlation energies obtained from different OEP
approaches and compare them with those from standard
correlated methods.
Although the correlation energies are defined differently in WFT and KS
theories these differences have been shown to be small, see for
example Refs.  \onlinecite{grabowski:2011:jcp,gros96}. In Table II we make use of this in comparing the WFT
MP2 and CCSD(T) correlation energies with those from the OEP
approaches investigated in the this work.

\begin{table*}
\caption{\label{tab2}Correlation energies (m$E_{\mathrm{h}}$) for different systems as obtained from several OEP2, 
OEP2-SOS, and conventional wave function methods. The last lines report the mean absolute error (MAE), 
and the mean absolute relative error (MARE) with respect to the CCSD(T) results 
In each line the best result among OEP2 
calculations is highlighted in bold.
The last two columns report the SOS coefficients from Eqs. 
(\ref{cOS1}) and (\ref{cOS2}).
 }
\begin{ruledtabular}
\begin{tabular}{lrrrrrcrrrcrr}
& \multicolumn{5}{c}{OEP2-}&&  \multicolumn{3}{c}{Wavefunction} && \\
\cline{2-6} \cline{8-10} \\
System & GL2  &  sc  &  SOSa  &  SOSb  &  SOSh  &&  SOS-MP2  &  MP2  &  CCSD(T) &&$c_{os}^{a}$  & $c_{os}^{b}$  \\ 
\cline{1-10} \cline{12-13} \\
He & -46.24 & -35.33 & -45.17 & -35.42 & {\bf -46.05} && -45.86 & -35.28 & -40.85             && 0.997 & 0.767\\ 
Be & nc & -70.76 & -96.81 & -71.49 & {\bf -92.94}          && -88.45 & -69.45 & -89.35        && 0.766 & 0.590 \\ 
Ne & -457.82 & -352.75 & {\bf -353.31} & -269.41 & -350.23 && -341.71 & -345.84 & -353.35     && 1.050 & 0.808 \\ 
Mg & -277.20 & -209.39 & {\bf -212.40} & -162.42 & -211.15 && -207.71 & -202.27 & -214.72
&& 0.977 & 0.751 \\ 
Ar & -776.35 & {\bf -625.70} & -594.59 & -456.85 & -593.91 && -591.56 & -617.09 & -641.00
&& 1.053 & 0.810 \\ 
H$_2$ & -48.55 & -32.20 & -42.31 & -32.40 & {\bf -42.11} && -41.59 & -31.99 & -39.72
&& 0.874 & 0.672 \\ 
He$_2$ & -92.47 & -70.66 & -92.19 & -70.84 & {\bf -92.09} && -91.73 & -70.56 & -81.70
&& 0.997 & 0.767\\ 
HF & -464.76 & -335.67 & -339.34 & -258.23 & {\bf -335.70} && -325.01 & -327.57 & -337.29
&& 0.995 & 0.766 \\ 
CO & -863.12 & {\bf -479.81} & -518.55 & -383.87 & -499.03 && -456.78 & -452.30 & -475.78
&& 0.912 & 0.701\\ 
H$_2$O & -466.33 & -322.82 & {\bf -331.08} & -251.73 & -327.25 && -316.95 & -314.26 & -329.17
&& 0.958 & 0.737\\ 
HCl & -516.48 & {\bf -384.38} & -374.01 & -286.92 & -373.00 && -369.90 & -375.10 & -402.17
&& 0.975 & 0.750 \\ 
Cl$_2$ & -1025.12 & {\bf -748.57} & -721.70 & -552.04 & -717.65 && -705.36 & -722.49 & -770.59
&& 0.967 & 0.744 \\ 
N$_2$ & -866.33 & {\bf -496.70} & -523.82 & -391.91 & -509.49 && -474.28 & -471.54 & -491.60
&& 0.885  & 0.681\\ 
Ne$_2$ & -915.95 & -705.77 & {\bf -706.77} & -538.91 & -700.58 && -683.53 & -691.85 & -706.91
&& 1.050 & 0.808\\ 
NH$_3$ & -421.59 & -290.96 & {\bf -303.86} & -231.85 & -301.41 && -294.18 & -283.97 & -305.65
 && 0.941 & 0.724 \\ 
C$_2$H$_6$ & -693.66 & -477.65 & {\bf -505.89} & -386.78 & -502.82 && -493.57 & -462.86 & -512.55
 && 0.928 & 0.714\\ 
&&&&&&&&&\\
MAE & 148.59$^a$ & {\bf 10.20} & 15.38 & 92.90 & 14.59 && 19.81 & 19.87 &&  \\ 
MARE & 35.32\%$^a$ & 4.96\% & 4.68\% & 22.53\% & {\bf 4.62}\% && 5.71\% & 7.11\%
&&  \\ 
\end{tabular}\\
\end{ruledtabular}
\begin{flushleft} 
nc -- not converged. \\
$^a$ without Be.
\end{flushleft} 
\end{table*}
Inspection of the table confirms the well known
features of the OEP-GL2 method, which yields 
much too negative
correlation energies
\cite{mori-sanchez:2005:oeppt2,engel:2005:oeppt2},
and of the OEP2-sc method, which is instead very accurate in this context
\cite{bartlett:2005:abinit2,grabowski:2007:ccpt2,grabowski:2011:jcp}. Concerning the OEP2-SOS approaches, we note that,
as expected, the OEP2-SOSh method is the most accurate
for the correlation energy, yielding a mean
absolute error (MAE), calculated with respect to CCSD(T) results, of only
14.59 m$E_{\mathrm{h}}$ and therefore 
slightly better than that of 
OEP2-SOSa (15.38 m$E_{\mathrm{h}}$), close to  
the one of the OEP2-sc method and better than both
MP2 and SOS-MP2. On the other hand, as anticipated by the
analysis of the previous subsection, an overestimation
of the correlation energy is given by the OEP2-SOSb method.
Nevertheless, this is almost completely removed when
the hybrid OEP2-SOSh method is considered
giving also the best  
mean absolute relative error (4.6\%).
In the last two columns of Table \ref{tab2}, the SOS coefficients from Eqs. 
(\ref{cOS1}) and (\ref{cOS2}) are reported
 (note that $c_{\mathrm{OS}}^{\mathrm{a}}=1.3 c_{\mathrm{OS}}^{\mathrm{b}}$).
We can see that the $c_{\mathrm{OS}}^{\mathrm{a}}$ coefficient is quite system dependent
with variation in the range $0.766-1.053$, meaning that a single SOS coefficient, as is used in
SOS-MP2 calculations \cite{WCMS12,grimme03}, cannot yield very accurate results for all systems.

In Table \ref{tab3} we report the integrated density differences (IDD, see
 Eq. (\ref{eidd})) corresponding to different OEP2 calculations.
\begin{table*}
\caption{\label{tab3} Integrated density difference (IDD; Eq. (\ref{eidd})) in units of $10^{-2}$  corresponding to various OEP2 methods. 
MP2 results are also shown for comparison. The last lines report the mean absolute error (MAE)  and the MAE weighted
for the number of electrons (MAE/$N_e$) 
In each line the best result for OEP2 or OEP2-SOS calculations is highlighted in bold.}
\begin{ruledtabular}
\begin{tabular}{lrrrrr}
System & OEP-GL2  & OEP2-sc  & OEP2-SOSa  & OEP2-SOSb/h  & MP2  \\ 
\hline
He  &       0.45  & 0.44 & 0.42 & {\bf 0.19} & 0.43 \\ 
Be  &         nc  & 3.13 & 4.59 & {\bf 1.73} & 4.30 \\ 
Ne  &       14.74 & {\bf 2.66} & 6.70 &  2.73 & 1.53 \\ 
Mg  &        5.60 & {\bf 2.20} & 3.54 & 4.14 & 1.80 \\ 
Ar  &        5.88 & {\bf 0.75} & 2.86 & 2.59 & 0.36 \\ 
H$_2$ &  0.44 & 0.38 & {\bf 0.24} & 0.34 & 0.45 \\ 
He$_2$  & 0.35 & {\bf 0.32} & 0.35 & 0.57 & 0.24 \\ 
HF  &     6.36 & 1.65 & 3.23 & {\bf 1.56} & 1.79 \\ 
CO  &     21.37 & {\bf 2.34} & 8.10 & 3.70 & 1.75 \\ 
Cl$_2$    & 13.93 & {\bf 4.11} & 8.53 & 5.41 & 2.19 \\ 
N$_2$     & 19.30 & {\bf 1.75} & 6.99 & 3.25 &1.27 \\ 
Ne$_2$    & 6.69 & {\bf 1.73} & 3.35 & 4.29 & 1.13 \\ 
HCl       & 4.25 & {\bf 2.28} & 3.18 & 2.45 & 1.27 \\ 
H$_2$O & 5.64 & 1.47 & 2.47 & \textbf{1.01} & 1.50 \\ 
NH$_3$ & 3.94 & 1.01 & 1.53 & \textbf{0.60} & 1.03 \\ 
C$_2$H$_6$ & 3.49 & \textbf{0.51} & 1.70 & 0.95 & 0.48 \\ 
\\
MAE & 7.50$^{a}$ & \textbf{1.67} & 3.61 & 2.22 & 1.35 \\ 
MAE/\textit{N$_e$} & 0.56$^{a}$ & \textbf{0.18} & 0.32 & 0.19 & 0.17 \\
\end{tabular}
\end{ruledtabular}
\begin{flushleft} 
nc -- not converged. \\
$^a$ without Be.
\end{flushleft} 
\end{table*}
The reported IDDs show that the best correlated  densities are found by the
OEP2-sc approach, while quite large errors are obtained by the
original OEP-GL2 method. We note that the OEP2-sc results are, for
most systems, in line with those of the MP2 calculations, showing
the reasonable accuracy of this functional.
A similar accuracy to that of OEP2-sc 
is observed also for the OEP2-SOSb and
OEP2-SOSh results, which yield an MAE of 0.022 and an MAE per electron
of 0.002 (note that the two 
methods coincide for this property) to be compared with the
values of 0.017 and 0.002 for OEP2-sc. 
As expected  slightly larger IDD values are obtained for OEP2-SOSa calculations.
Nevertheless, we note that the performance of the 
OEP2-SOSa method is good, being about a factor of two 
better than the OEP-GL2 method.

Similar trends to those observed for the IDDs are found when the 
HOMO-LUMO gaps are considered. This is shown in
Table \ref{tab4} where we consider the HOMO-LUMO gap computed by 
various OEP2 and OEP2-SOS methods and compare them to the
benchmark data obtained by direct inversion of the relaxed
densities calculated from several {\it ab initio} methods.
\begin{table*}
\caption{\label{tab4} HOMO-LUMO energy gap (eV) for different systems as obtained from several OEP2 and 
OEP2-SOS approaches. These are compared with those from the KS inversion procedure applied to conventional 
correlated methods. The last lines report the mean absolute error (MAE), and the mean absolute relative error 
(MARE) with respect to the CCSD(T) results.
In each line the best result for OEP or OEP2-SOS calculations is highlighted in bold.}
\begin{ruledtabular}
\begin{tabular}{lrrrrrcrr}
& &\multicolumn{4}{c}{OEP2-} && \multicolumn{2}{c}{Inverse-KS} \\
\cline{3-6} \cline{8-9}
System & OEPx & GL2 & sc & SOSa & SOSb/h && MP2 & CCSD(T) \\
\hline
He & 21.60 & 20.95 & 21.32 & 20.96 & {\bf 21.11}     && 21.33 & 21.21 \\ 
Be & 3.57 & nc & {\bf 3.63} & 3.33 & 3.50            && 3.63 & 3.61 \\ 
Ne & 18.48 & 14.12 & 16.45 & 15.76 & {\bf 16.46}     && 16.81 & 17.00 \\ 
Mg & 3.18 & 3.40 & {\bf 3.33} & 3.37 & {\bf 3.33}    && 3.33 & 3.36 \\ 
Ar & 11.80 & 10.95 & {\bf 11.43} & 11.22 & 11.37     && 11.45 & 11.51 \\ 
H$_2$ & 12.09 & 12.03 & {\bf 12.13} & 12.04 & 12.05 && 12.18 & 12.14 \\ 
He$_2$ & 21.28 & {\bf 20.64} & 21.02 & {\bf 20.64} & 20.80 && 20.68 & 20.56 \\ 
HF & 11.36 & 7.80 & {\bf 9.84} & 9.18 & 9.74         && 10.13 & 10.30 \\ 
CO & 7.77 & 5.87 & {\bf 7.22} & 6.79 & 7.07         && 7.32 & 7.29 \\ 
H$_2$O & 8.44 & 5.99 & {\bf 7.49} & 7.04 & 7.41    && 7.59 & 7.75 \\ 
HCl & 7.82 & 7.10 & {\bf 7.52} & 7.28 & 7.41      && 7.60 & 7.55 \\ 
Cl$_2$ & 3.90 & 2.65 & {\bf 3.35} & 2.98 & 3.21   && 3.32 & 3.29 \\ 
N$_2$ & 9.21 & 6.73 & {\bf 8.37} & 7.86 & 8.22      && 8.46 & 8.55 \\ 
Ne$_2$ & 17.84 & 13.49 & 15.75 & 15.06 & {\bf 15.76} && 16.05 & 16.23 \\ 
NH$_3$ & 6.97 & 5.30 & {\bf 6.35} & 6.00 & 6.25 && 6.41 & 6.54 \\ 
C$_2$H$_6$ & 9.21 & 8.24 & 8.85 & 8.75 & {\bf 8.87} && 8.82 & 8.95 \\ 
&&&&&&&\\
MAE &  0.52     & 1.15$^a$ & {\bf 0.21} & 0.50 & 0.24 && 0.10 &  \\ 
MARE & 6.49$\%$ & 12.16\%$^a$ & {\bf 1.86}\% & 5.20\% & 2.48\% && 1.02\% &  \\ 
\end{tabular}
\end{ruledtabular}
\begin{flushleft} 
nc -- not converged. \\
$^a$ without Be.
\end{flushleft} 
\end{table*}
The data in the table show that the best performance is
given by the OEP2-sc method, which yields an MAE of 0.21 eV,
close to the estimated accuracy of the reference data ($\sim 0.2$ eV).
The OEP2-sc method displays a strong improvement with respect
to the OEP-GL2 approach, which instead gives
a systematic underestimation of
the HOMO-LUMO energy gap. A significant improvement
with respect to OEP-GL2 is also obtained by considering
various scaled-opposite-spin OEP methods.
 In fact, a reduction of both $c_{\mathrm{SS}}$ and $c_{\mathrm{OS}}$
tends to increase the HOMO-LUMO gap, due to the
reduced weight of the correlation contributions in the 
XC potential.
This shows that the scaling of the 
opposite-spin correlation is an effective way to
improve the description of the correlation potential.
Accurate results are given in particular by the
OEP2-SOSb method (note that for this property
OEP2-SOSb and OEP2-SOSh coincide by definition),
which yields an MAE of 0.24 eV and an MARE of 2.5\%.
Slightly larger errors are obtained with the 
OEP2-SOSa method, which displays a moderate tendency
to underestimate the HOMO-LUMO energy gap.

\begin{figure}
\begin{center}
\includegraphics[width=2\columnwidth,angle=270]{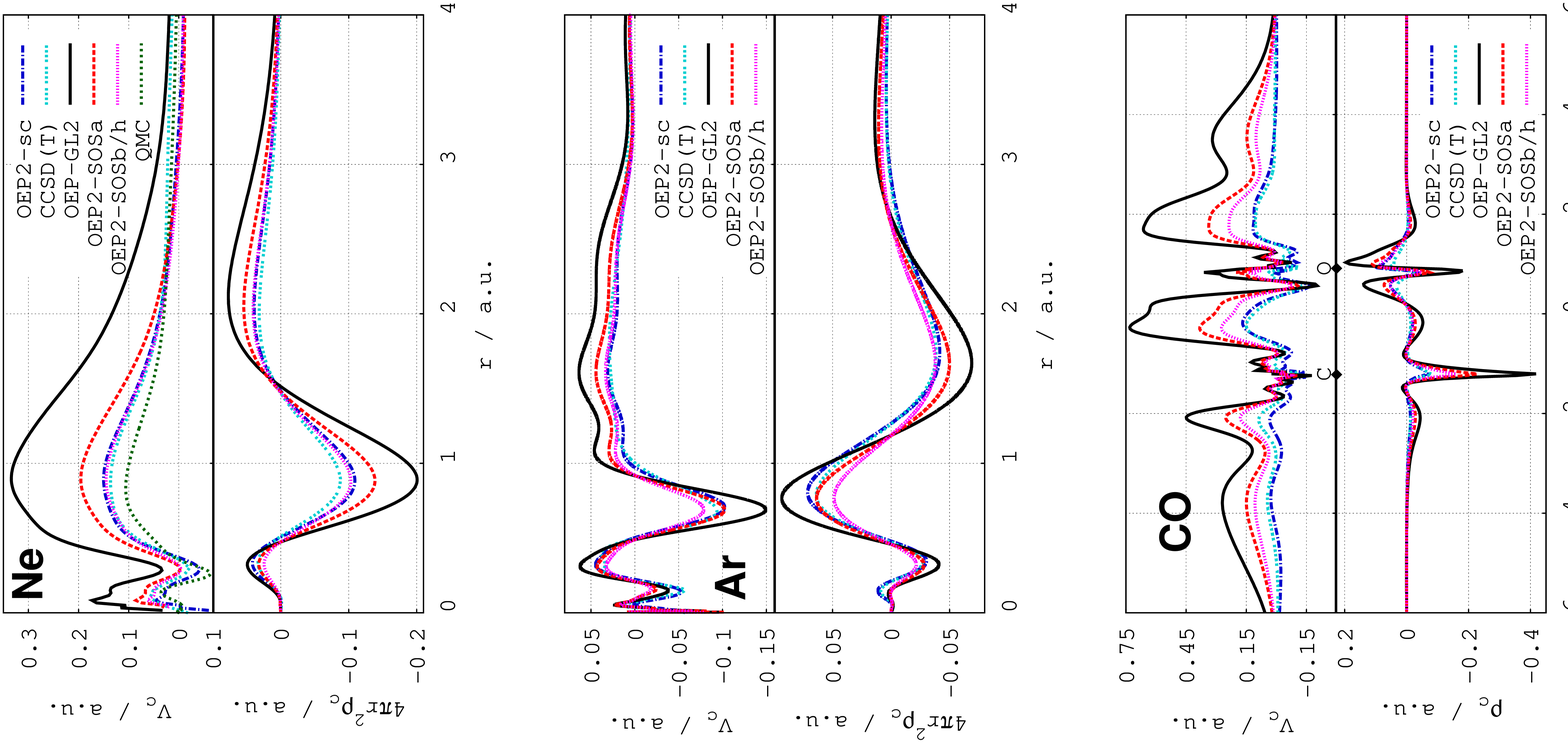}
\caption{\label{sos-H2O_Ne} Correlation potential and density contribution, $\rho_{\mathrm{c}}$, for the
Ne atom (upper panel), Ar atom (middle panel) and the CO molecule (lower panel), as obtained from
different OEP methods as well as benchmark correlated methods and corresponding
inverted KS calculations. 
}
\end{center}
\end{figure}

To conclude this section we consider in Figure~\ref{sos-H2O_Ne} a direct comparison of the
correlation potentials and density differences
for several methods. The plots are reported
for the Ne, Ar and CO systems and are representative
of a general trend.

Inspection of the figure confirms the observations
made above on the behaviour of different functionals.
All plots show in fact, the significant overestimation
of the correlation potential by the OEP-GL2 method,
especially in the valence regions, and the
important improvement obtained by considering
OEP2-SOS methods. This improvement is, of course,
more pronounced for the OEP2-SOSb potential,
which is qualitatively comparable with the accurate
OEP2-sc one. We observe a systematic improvement of the correlation
potentials going from OEP-GL2, through OEP2-SOSa to the OEP2-SOSb/h method
which gives the best correlation potentials as compared to the CCSD(T)
results.
In fact, all methods
provide also a qualitatively similar description of the
correlated density. Moreover, 
the OEP2-SOSb and OEP2-sc plots are very close to the CCSD(T) one.
\section{Conclusions}\label{sec:Conc}
We have investigated the performance of OEP2-SCS functionals
to elucidate the role of the different 
spin-resolved singly-- and doubly--excited contributions,
as well as the importance of the independent scaling of such terms.
We found that the most important contribution
to the description of correlation comes from the doubly
excited terms. Moreover, 
the same- and opposite-spin contributions to the dominant doubly excited terms 
 were found to
display a significant proportionality, so that
for both the correlation energy and the density-based
properties it is possible to individuate a 
continuous family of OEP2-SCS
functionals of high accuracy, differing only in the coefficients used for the
the same- and opposite-spin terms.
This finding is important because it
provides a rationale for the well known
scaled-spin-component wave-function approaches, 
 which up to now have been based on empirical observations
focusing only on the correlation energy.
In addition, the flexibility in the choice of the
``best''  $c_{\mathrm{OS}}$ and $c_{\mathrm{SS}}$ parameters, clearly
suggests that accurate functionals
can be defined also in the
realm of the scaled-opposite-spin correlation,
which is particularly attractive due to its favourable computational 
cost.   

We considered three possible variants of OEP2-SOS functionals,
based on different
approaches for determining the $c_{\mathrm{OS}}$ parameter, and
assessed their performance on a representative set of different
systems and properties. Amongst the methods considered, 
the OEP2-SOSa approach gave the most accurate correlation energies,
whereas the OEP2-SOSb method displayed higher accuracy for the
description of the electron density, correlation potentials,
HOMO-LUMO gaps and orbital energies.
Our findings are consistent with similar observations
  \cite{Kronik2014}  made for
some standard DFT methods which, depending on the choice of the
parameters used for defining the XC functional, are either
accurate for binding energies or orbital energies, but rarely
for both properties at the same time.

Thus, the OEP2-SOSh approach, using a post-SCF rescaling of the 
correlation energy was constructed as a robust method for the description
of correlation energies, correlated densities, and single particle properties;
displaying a similar performance to the OEP2-sc method, which is
one of the most advanced OEP 
second-order correlated
approaches currently available. 

An interesting point in favour of the OEP2-SOS family of functionals
is that they display not only improved correlation energies, correlation
potentials, HOMO-LUMO gaps and correlated densities but also
improved stability in comparison with the OEP-GL2 method. The latter
is unbounded from below and can exhibit variational collapse for unphysical
exchange--correlation 
potentials~\cite{rohr-baerends:2006:oephe,engel:2005:oeppt2}. 
However, these potentials do not correspond to KS solutions and
can be easily excluded by minimal regularization (such as the truncated SVD
approach used in the present work). Nonetheless there can remain difficult
cases where the calculations do diverge even with regularization due to the
breakdown of the perturbation theory. A prototypical case is the Be atom,
for which OEP-GL2 diverges whereas the OEP2-SOS approaches in the present
work remain stable for the parameters calculated in SOS-OEP2a (
$c_{\mathrm{OS}}=0.766$) and SOS-OEP2b ( $c_{\mathrm{OS}}=0.590$)
ansatzes. Actually the SOS-OEP2 method converges for the whole range of
$c_{\mathrm{OS}}$
parameter from  $c_{\mathrm{OS}}=0.0$ up to  $c_{\mathrm{OS}}=0.770$. 
 
In conclusion, the present work shows that scaled-spin-component,
and especially scaled-opposite-spin, methods are promising
approaches. This applies not only in the context of \textit{ab initio} correlated methods,
where they are currently increasingly employed in many application
and development works, but also in the field of DFT. The availability of OEP2-SOS functionals,
which exhibit a favourable computational cost (with respect
to similar 
  correlated OEP
 approaches) and good accuracy, should allow
for applications to larger and more complex systems.
This will provide a valuable tool for practical studies
and, more importantly, a deeper understanding of the
behaviour of DFT correlation. This information is essential for the further
development of density--functional approximations 
and may be of utility for improving the quality lower cost of semi-local
forms.

\section{Acknowledgments}
This work was partially supported by
the National Science Center under
Grants No. DEC-2012/05/N/ST4/02079, DEC-2013/08/T/ST4/00032, and
DEC-2013/11/B/ST4/00771, 
and the European Research Council
(ERC) Starting Grant FP7 Project DEDOM, Grant No. 207441. A. M. T. gratefully
acknowledges support via the Royal Society University Research Fellowship
scheme.


\end{document}